# A Unified Architecture for Doubly Fed Induction Generator Wind Turbines using a Parallel Grid Side Rectifier and Series Grid Side Converter

Patrick S. Flannery  Giri Venkataramanan

University of Wisconsin – Madison

1535 Engineering Hall

1415 Engineering Drive

Madison, WI 53706, USA

*psflannery@wisc.edu*  *giri@engr.wisc.edu*

***Abstract*— With steadily increasing wind turbine penetration, regulatory standards for grid interconnection are evolving to require that wind generation systems ride-through disturbances such as faults and support the grid during such events. Conventional modifications to the doubly fed induction generation (DFIG) architecture for providing ride-through result in limited control of the turbine shaft and grid current during fault events. A DFIG architecture in which the grid side converter is connected in series as opposed to parallel with the grid connection has shown improved low voltage ride through but poor power processing capabilities. In this paper a unified DFIG wind turbine architecture which employs a parallel grid side rectifier and series grid side converter is presented. The combination of these two converters enables unencumbered power processing and improved voltage disturbance ride through. A dynamic model and control structure for this unified architecture is developed. The operation of the system is illustrated using computer simulations.**

## I. INTRODUCTION

As the penetration of large scale wind turbines into the electric power grid continues to increase, electric system operators are placing greater demands on wind turbine power plants. One of the most challenging new interconnection demands for the DFIG architecture is its ability to ride through a short-term low or zero voltage event at the point of common coupling (PCC), resulting from a fault on the grid. During extreme voltage sags high per unit currents and shaft torque pulsations occur unless mitigating measures are taken.

Low voltage ride through requirements were first proposed by German electric transmission operators E.ON and VE-T in 2003 [1]. A modified voltage ride through requirement was recently adopted in the U.S. via FERC orders 661 and 661A. The U.S. ride through requirement stipulates that the wind turbine must remain connected to the grid and provide fault clearing current in the event that the voltage at the high side of the step up transformer to the transmission system drops to zero volts for a maximum of 9 cycles, as the result of a three phase fault [2].

In a conventional DFIG wind turbine the machine stator windings are connected to the grid PCC via collection and/or transmission transformers and excited at the grid frequency. The rotor windings of the DFIG are connected to an AC/DC converter commonly referred to as the machine side converter (MSC). The ac side of a second DC/AC converter, commonly referred to as the grid side converter (GSC), is connected in parallel with the machine stator windings and PCC. A transformer enables voltage and current compatibility between the GSC and the PCC. The ac ports of each converter are connected to form a dc link, enabling power flow between them. Both the MSC and GSC are controlled with inner current loops, and outer loops to control torque and reactive power production [3].

Severe voltage sags place significant electrical stress on the IGBT and diodes of the MSC and the mechanical stress on the rotor gearbox. When the stator voltage changes abruptly, the stator flux responds in the manner of a lightly damped oscillatory system [4]. This oscillatory stator flux couples to the rotor circuit and produces oscillations in the electromagnetic torque. In the event of deeper voltage sags and larger stator flux oscillations, MSC voltage limit is exceeded resulting in loss of rotor current regulation. The uncontrolled rotor currents typically exceed the semiconductor device ratings and result in damage the MSC. In addition, this also precipitates very large stator currents and transient spikes in shaft torque. [5].

In order to manage these problems, modifications to the conventional DFIG architecture for ride-through have been proposed in the past. The first of these is a rotor crowbar clamp, realized with three pairs of back-to-back SCRs and a three phase resistor network in parallel with the MSC. In the event of an extreme voltage sag, this clamp diverts current from the MSC. Unfortunately, large currents still flow through the rotor windings of the DFIG, resulting in large torque spikes. An alternate ride through modification uses set of stator side SCR switches to enable brief disconnection of the stator windings inception of the sag event [6]. The MSC synchronizes the flux in the machine with the new voltage level and the stator windings are reconnected to the PCC for the remainder of the sag event. The resulting torque pulsations are reduced somewhat, but this topology still requires the rotor crowbar for emergency protection of the MSC.

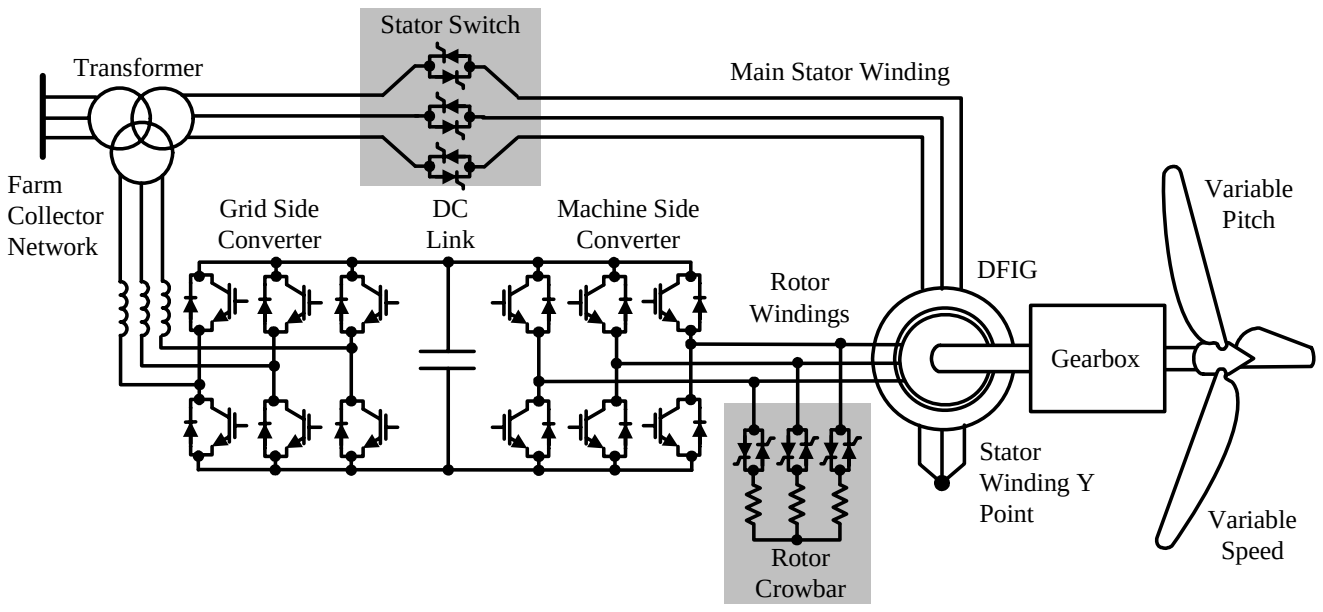

Fig. 1. Schematic of conventional DFIG wind turbine architecture with rotor crowbar and stator switch modifications to improve low voltage ride-through.

As an alternative to these traditional modifications, an arrangement in which the grid side converter was connected in series with the stator windings of the DFIG was proposed in [7], but it properties and limitations were not studied in depth to develop a definitive solution. Further exploration of the series grid side converter DFIG architecture in [8] revealed short comings in the power processing capabilities and unified architecture that is able to successfully overcome the limitations was identified. The new unified DFIG architecture in which the series grid side converter is partnered with a parallel grid side rectifier appears as an attractive option for both DFIG wind turbine power processing and improved voltage sag ride through.

The focus of this paper is to develop a dynamic model and control structure for this unified DFIG architecture and characterize its steady state and dynamic response properties. This paper is organized in the following manner: Section II presents a dynamic model and block diagrams of the unified DFIG wind turbine architecture; an overall control strategy is developed in Section III, followed by detailed design of each local controller in Section IV. Dynamic response results developed using computer simulations are presented in Section V to verify successful operation of the system to both PCC voltage sags and wind speed variations.

## II. Unified DFIG Architecture System & Model

### A. System description

Mechanically the unified DFIG architecture wind turbine is similar to conventional DFIG wind turbines. The rotor speed is allowed to vary in a limited range around the synchronous speed allowing optimal aerodynamic energy capture. Blade pitch is adjustable to feather the blades and throttle back energy capture. A multistage gearbox increases the turbine shaft speed for compatibility with the DFIG.

The simplified electrical interconnection of the proposed DFIG architecture employing both series grid side converter and parallel grid side rectifier is presented in Fig. 2. As in a conventional DFIG, the main stator windings of the machine are connected, via transformer, to the line excitation. The rotor windings of the machine are accessed via slip rings and connected to a voltage source converter referred to as the machine side converter (MSC). The dc terminals of the MSC are connected to a full bridge diode rectifier referred to as the parallel grid side rectifier (PGSR) which is excited from the line via transformer. The dc link is also connected to a second voltage source converter called the series grid side converter (SGSC). The three phase terminals of the SGSC are connected in series with the three phase terminals of the stator circuit and the line. Thus, the injected stator power also passes through the SGSC, providing the active mechanism necessary for providing ride through during grid disturbances in an effective manner. Individual wind turbines are further grouped together in a local electrical power collection network, also known as a wind farm, for connection to the grid.

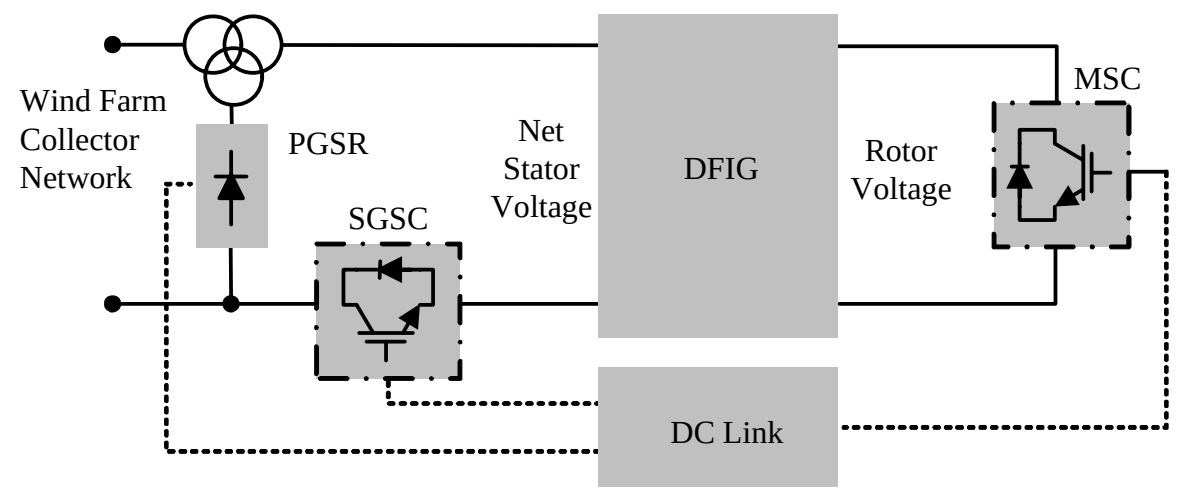

Fig. 2. Simplified electrical schematic of unified DFIG architecture with parallel grid side rectifier (PGSR) and series grid side converter (SGSC)

A detailed schematic of the three phase power converters, stator and rotor windings of the DFIG, and the interconnection transformer is shown in Fig. 3. As may be observed from the figure, the Y point of the stator terminals DFIG are opened and the three stator return terminals of the machine and individually connected to the three phase legs of the SGSC. The three phase ac voltage output of the SGSC adds to the voltage from the farm collection network to yield the net stator voltage. As an alternative realization, the DFIG stator Y point may be maintained closed and a 3 phase series injection transformer can be used to connect the GSC between the collection network voltage and main stator windings, such as in a dynamic voltage restorer [9]

### B. Electrical model

The electrical model for the system is developed using dynamic phasors or complex vectors in the synchronously rotating D-Q reference frame [10]. An illustration of the axes conventions appears in Fig. 4. The default convention assumed here aligns the Q-axis with the positive real axis and the D-axis with the negative imaginary axis, and the complex vector $\underline{z} = z_Q - jz_D$. However, in certain instances it is convenient to locate the real and imaginary axes aligned with a particular complex vector, for instance $\underline{\alpha}$, in which case the axes are designated $Q^{\alpha}$ and $D^{\alpha}$ respectively, and the real and imaginary components with respect to the $\underline{\alpha}$ reference are designated $z_Q^{\alpha}$ and $z_D^{\alpha}$ a respectively.

The following simplifying assumptions are made in the development of the model:

- The iron losses, mechanical and power converter losses are negligible.

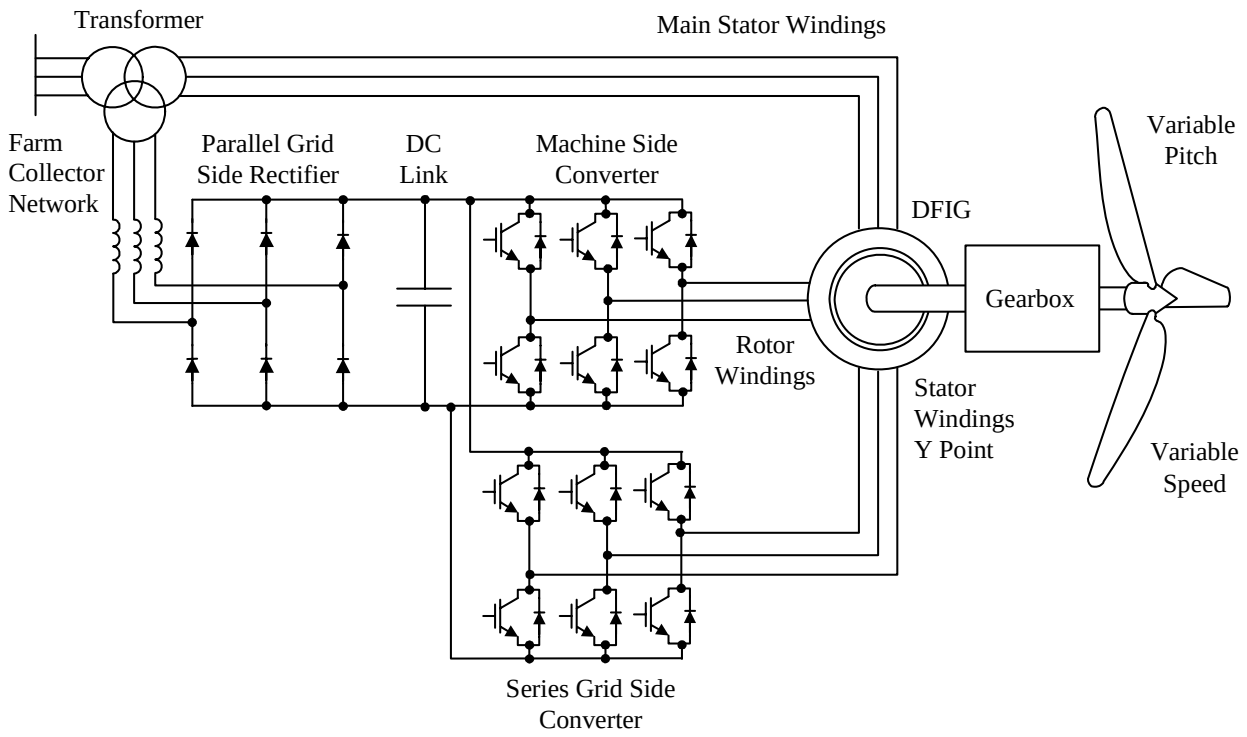

Fig. 3. Schematic of unified DFIG architecture with parallel grid side rectifier (PGSR) and series grid side converter (SGSC)

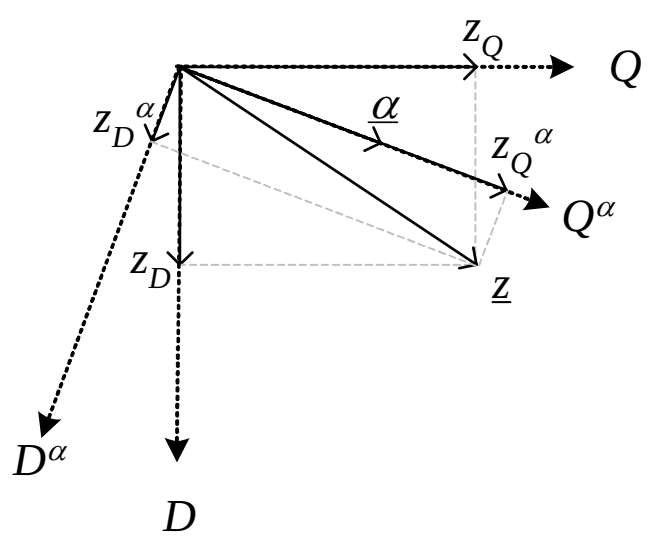

Fig. 4. Conventions and notation of complex vectors and reference axes

- The magnetic circuit of the machine can be represented by a linear model.
- The entire mechanical system can be modeled using a lumped inertia parameter referred to the electrical angle and velocity variables of the induction generator.
- The power converters can be modeled using its state-space averaged representation to represent their low frequency dynamics.
- The wind farm collection network referred to the secondary network is electrically stiff due to the relatively large underground collection cable capacitance.

An equivalent Γ circuit model for the DFIG is used to develop an orthogonal stator flux and rotor current oriented scheme for torque control [11]. The system equivalent circuit model under these assumptions is shown in Fig. 4.

The complex vector dynamic equations for the electrical system can be developed by applying Kirchoff's voltage law at the stator and rotor circuit loops as:

$$\frac{d\underline{\lambda}_s}{dt} = \underline{v}_f + \underline{m}_i v_{dc} - \underline{i}_s R_s - j\omega_e \underline{\lambda}_s \tag{1}$$

$$\frac{d\underline{\lambda}_R}{dt} = \underline{m}_R v_{dc} - \underline{i}_R R_R - j\omega_{sl} \underline{\lambda}_R \tag{2}$$

Since PGSC is a passive network, its conduction state is determined by the state of the diode illustrated in Fig. 4, which conducts when the voltage $\underline{v}_c$ is greater than $\underline{m}_{re} v_{dc}$. The vector dynamics of the current $\underline{i}_{re}$ may be represented by

$$\frac{di_{re}}{dt} = u(\underline{v}_f - \underline{m}_{re} v_{dc}) \left( \frac{\underline{m}_{re} v_{dc} - \underline{v}_c}{L_{re}} - j\omega_e i_{re} \right) \tag{3}$$

where u(**.**) is the unit step function. The dynamics of the dc bus can be represented using Kirchoff's current law at the dc bus as

$$\frac{dv_{dc}}{dt} = \frac{-3}{2C_{dc}} (\underline{m}_i \underline{i}_s + \underline{m}_{re} \underline{i}_{re} + \underline{m}_R \underline{i}_R) . \tag{4}$$

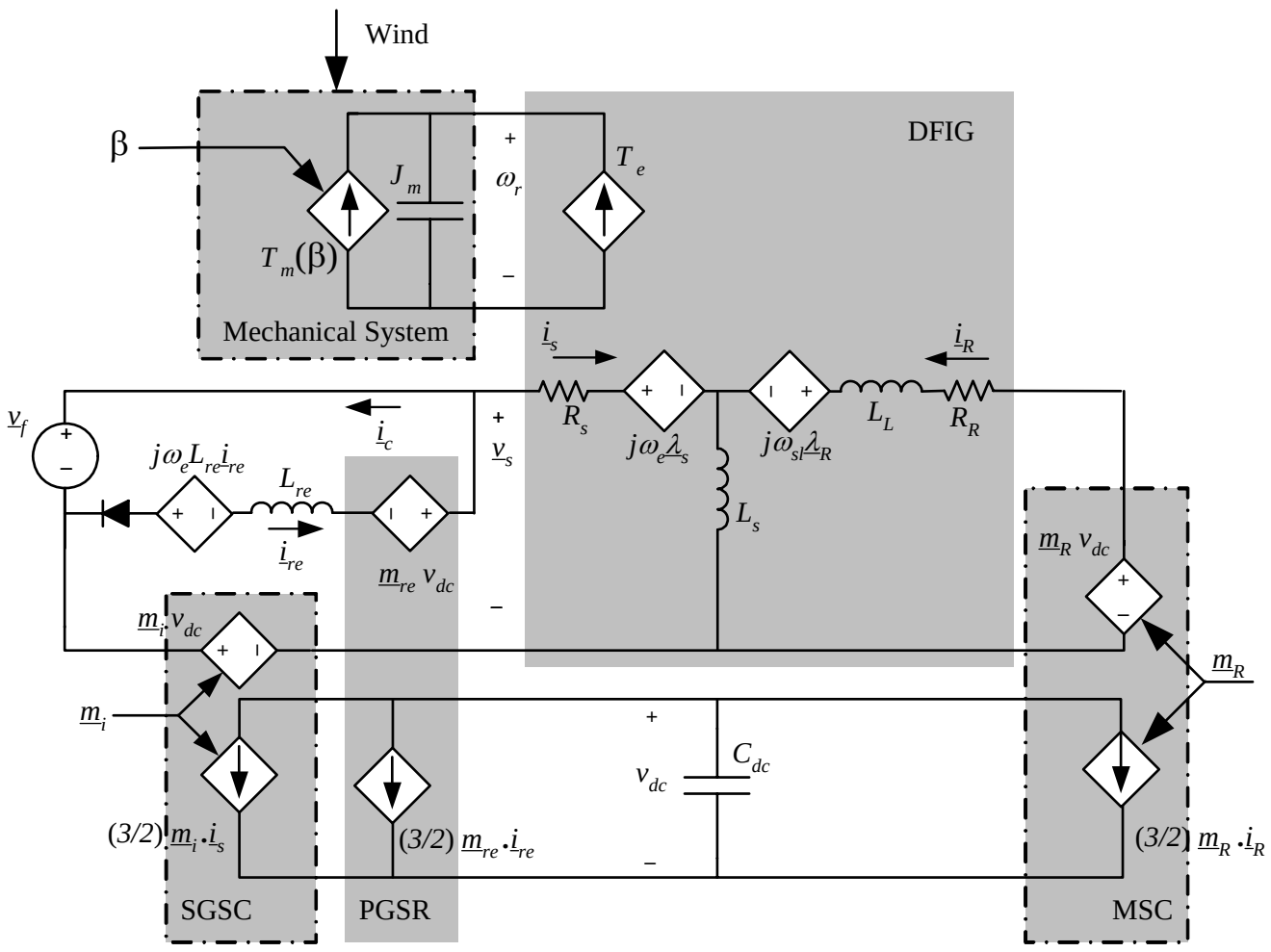

Fig. 5. Dynamic phasor circuit model of unified DFIG electrical architecture and mechanical system

The algebraic relationship between the rotor and stator currents, rotor and stator flux components and the electromagnetic torque can be expressed as

$$T_e = \frac{3P(\underline{\lambda}_s \times \underline{i}_R)}{4}$$
$$\underline{\lambda}_s = L_s(\underline{i}_s + \underline{i}_R) \ . \tag{5}$$
$$\underline{\lambda}_R = \underline{\lambda}_s + L_L \underline{i}_R$$

*C. Wind power and mechanical model*

The mechanical power generated at the wind turbine shaft is proportional to the coefficient of performance and the cube of the wind speed.

$$P_m = k_a C_P(\psi, \beta) u_w^{\ 3} . \tag{6}$$

The wind driven mechanical torque produced at the DFIG rotor shaft is the mechanical power divided by the rotor shaft speed, as in

$$T_m = \frac{k_a C_P(\psi, \beta) u_w^{\ 3} P K_{gb}}{2\omega_r} . \tag{7}$$

The coefficient of performance is maximized when the blade pitch is fixed at (or near) zero degrees, and the tip-speed-ratio is held at it optimal value, typically a constant for a given turbine and blade design. The analytical model of the coefficient of performance vs. blade pitch angle and tip-speed-ratio is given in [12]. The blade pitch actuators are assumed to have a flat low frequency dynamic response with a real pole at $\omega_\beta$ as described by

$$\frac{d\beta}{dt} = \omega_\beta (\beta^* - \beta) \tag{8}$$

The blade pitch actuators are also position and rate limited, with values of $0^o$ to $30^o$ degrees and +/- $10^o$/s respectively.

The rotor speed state equation can be expressed in terms of the total inertia and the sum of the electromagnetic and mechanically torques, as in

$$\frac{d\omega_r}{dt} = \frac{P}{2J_m}(T_e + T_m) \tag{9}$$

The relationships (1)-(9) together form the dynamic model for the proposed DFIG system. The energy inputs and outputs appear as $T_m$ and $\underline{v}_f$, while the control actuation is provided through the blade pitch, $\beta$; MSC modulation $\underline{m}_R$; and SGSC modulation $\underline{m}_i$. The steady state relationship between various parameters and variables are discussed in the following section to develop a suitable control structure that is proposed subsequently.

## III. Steady State Operation

### A. Power Flow

It is well known that the breakdown in the flow of mechanically generated power out of the DFIG stator and rotor terminals depend only the machine slip if the electrical losses in the machine are neglected. In this case, the net shaft power throughput of the machine ($P_m$) splits between the stator circuit ($P_s$) and rotor electrical circuit ($P_{MSC}$) following the relationships:

$$P_s \cong \frac{P_m}{(1-s)} \tag{10}$$

$$P_{MSC} \cong \frac{s\,P_m}{(1-s)} \tag{11}$$

Since there is no significant energy dissipation in the DC link, the power entering the dc bus from the MSC must leave through the PGSC and SGSC. Thus,

$$P_{SGSC} + P_{PGSR} \cong \frac{-s\,P_m}{(1-s)} \tag{12}$$

The power delivered to the SGSC, PGSR and stator windings sum to yield the total power delivered to the farm collector network, which is also equal to the mechanical shaft power in the case of zero losses. The power flow breakdown is illustrated graphically in Fig. 6.

At low and medium wind speeds, the blade pitch angle and tip-speed-ratio of the wind turbine shaft are typically controlled to maximize the mechanically captured power. The corresponding mechanical power is proportional to the cube of the rotor speed, expressed as

$$P_m = k_{opt}\omega_r^{\,3}\,. \tag{13}$$

At higher wind speeds, typically greater than 1.2 per unit the rotor speed is typically limited due to mechanical design considerations. A plot of the mechanical, stator and sum of SGSC and PGSR powers over a typical speed range is presented in Fig. 7. It may be observed that the SGSC and PGSC together process negative power (i.e. absorb power from the ac network) at subsynchronous speeds, and process positive power (i.e. deliver power into the ac network) at supersynchronous speeds.

### B. Operation with SGSC only

Having determined the power throughput across different converter elements of the proposed DFIG system, the next step is to determine the actual voltages and currents necessary to realize power flow. The case with the PGSR disabled (i.e. $P_{PGSR}$ = $\underline{i}_{re}$ =0) and SGSC alone is enabled is used to develop the functional properties, leading to the operational strategy for the overall system.

To develop a preliminary conceptual understanding of the power flow phenomena in the overall DFIG system, the stator circuit and the SGSC are all assumed to be unity power factor as a first step. These conditions are termed *universal unity power factor operation* (UUPF) and will be relaxed further to provide additional degrees of freedom at a later stage.

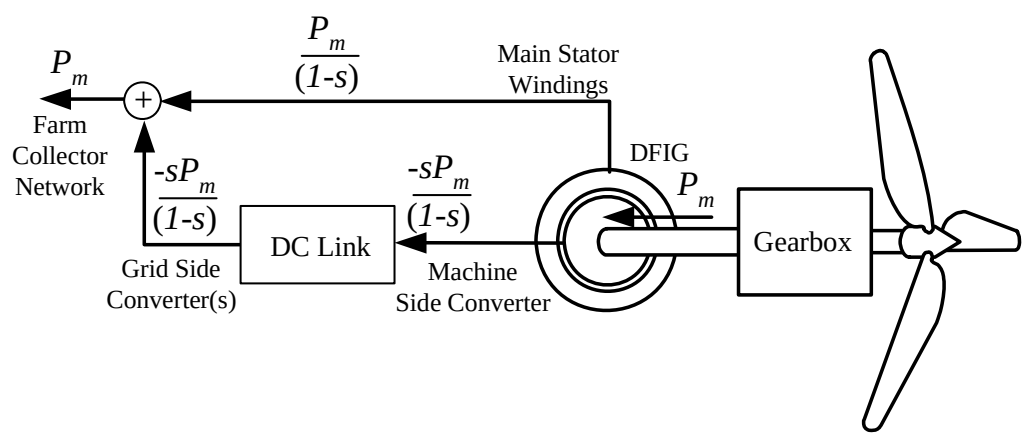

Fig. 6. Schematic representation of power flow breakdown in a DFIG wind turbine

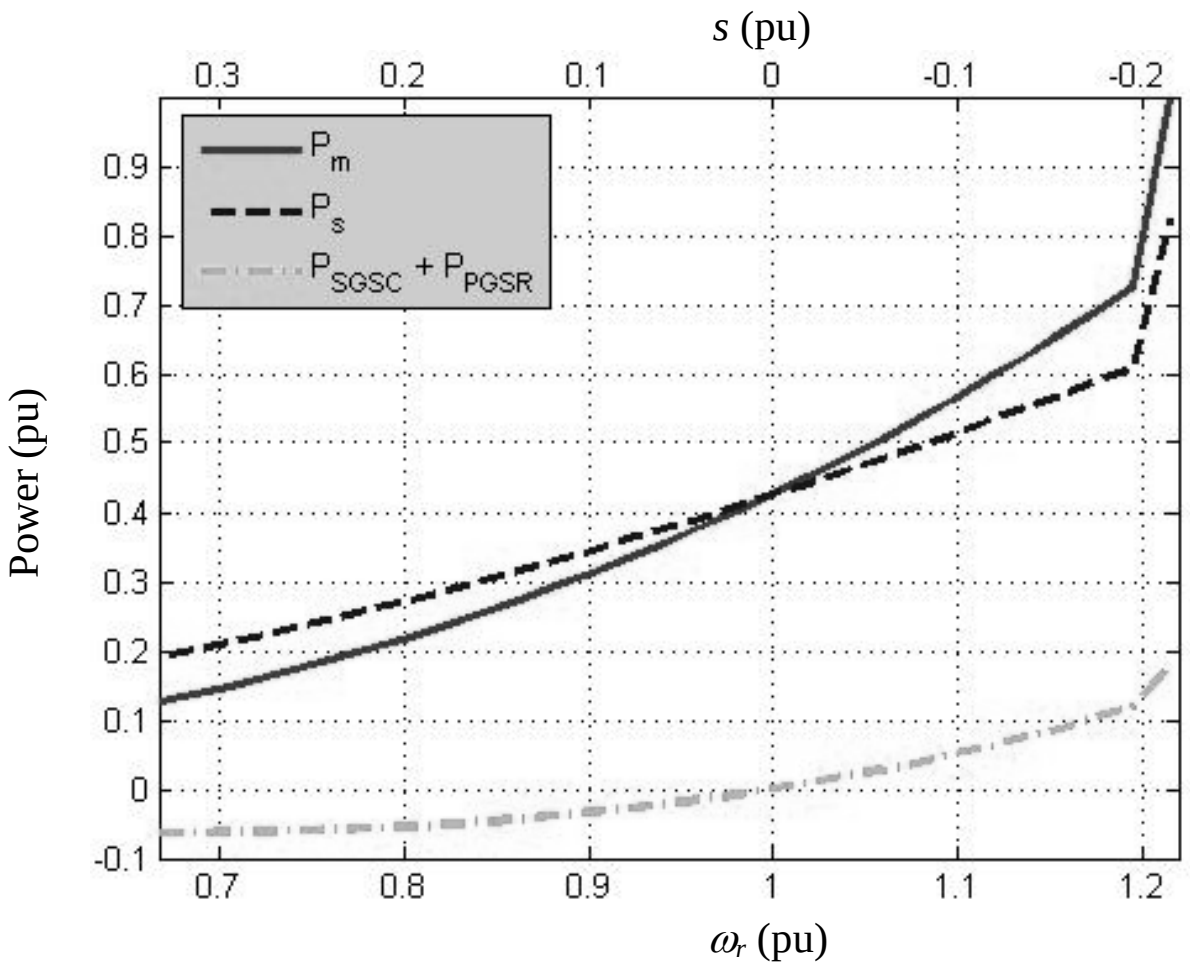

Fig. 7. Mechanical, stator and SGSC + PGSR power vs. rotor speed (bottom scale) and slip (top scale)

The stator terminal voltage is assumed to be the reference phasor for alignment of the positive Q-axis (i.e. $\underline{v}_s$ = $\underline{v}_{Qs}$). Since the overall DFIG system terminals, stator terminals and SGSC terminals are all assumed to operate at unity power factor, their corresponding currents ($\underline{i}_s = \underline{i}_f = \underline{i}_i = i_{Qf} = i_{Qi} = i_{Qs}$) and voltages may be related in terms of their respective throughput powers, as

$$v_{Qf} = \frac{-3}{2}\frac{P_m}{i_{Qf}};\, v_{Qs} = \frac{-3}{2}\frac{P_s}{i_{Qf}};\, v_{Qi} = \frac{-3}{2}\frac{P_{SGSC}}{i_{Qf}}\,. \tag{14}$$

Eliminating $i_{Qf}$ from (14), we obtain

$$v_{Qs} = \frac{P_s}{P_m}v_{Qf};\, v_{Qi} = \frac{P_{SGSC}}{P_m}v_{Qf} \tag{15}$$

Substituting for the ratio of powers in (15) from (10) and (12) respectively, the stator terminal voltage and the SGSC terminal voltages may be determined to be

$$v_{Qs} = \frac{1}{1-s}v_{Qf};\, v_{Qi} = \frac{-s}{1-s}v_{Qf} \tag{16}$$

Furthermore, from Fig. 5, the DFIG stator voltage may be expressed as the sum of the farm collector node and SGSC voltages as in

$$\underline{v}_s = \underline{v}_f + \underline{v}_i\,. \tag{17}$$

#### i) Supersynchronous operation

A phasor diagram of the stator current and various voltages under UUPF operation is shown in Fig. 8. All the stator circuit phasors are located along the real axis. It may be observed that $v_{Qs}$<$|v_f|$, and $v_{Qi}$<0. Therefore, operating the machine under UUPF approach leads to reduced stator voltage and stator flux. This leads to reduced utilization of machine capability at supersynchronous speeds.

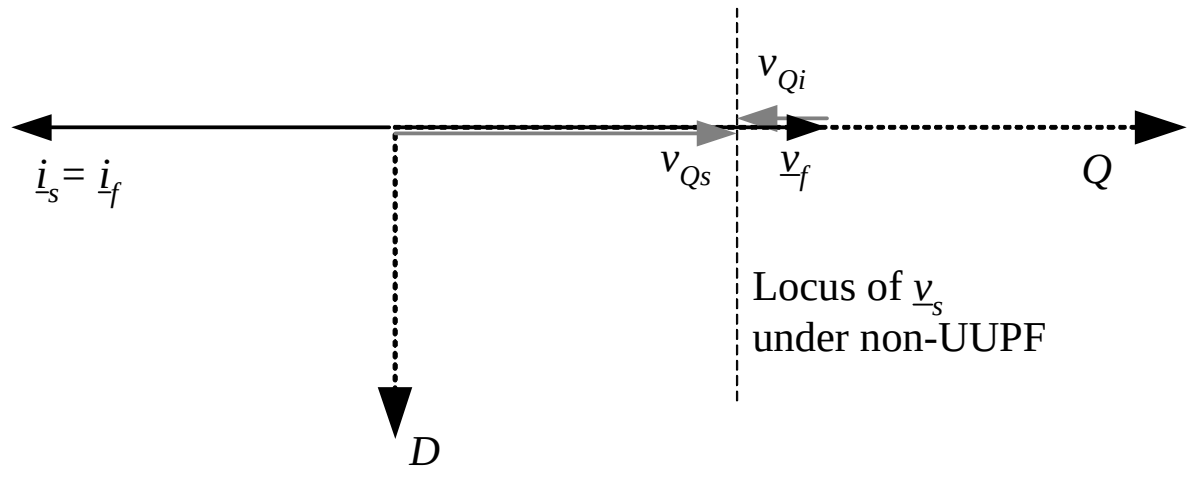

Fig. 8. Phasor diagrams of stator circuit voltages and currents for unified DFIG architecture with only SGSC at supersynchronous speeds

Since it would be typically desirable for the DFIG to operate at rated stator voltage (and flux) for full utilization of machine capability, the UUPF constraint will need to be relaxed. In this case, the phasors $\underline{v}_f$ and $\underline{i}_s$ remain on the real axis while the phasors $\underline{v}_i$ and $\underline{v}_s$ are free to move along the dotted line shown in Fig. 8. The voltage loci still follow the Kirchoff's voltage law expressed in (17). The Q-components of voltages still follow (16) to maintain the overall power transfer conditions, and the D-components of stator and SGSC voltages are identical. Thus, adequate D-axis voltage may be injected using the SGSC which will be balanced by an equal amount of D-axis voltage across the stator terminals. The total stator voltage is restored to 1 p.u. as illustrated in Fig. 9.

As may be observed from Fig. 9, both positive and negative values of D-axis SGSC voltage may be used to restore the stator voltage to unity. The phasors corresponding to a positive D-axis voltage are shown in solid lines, with a '+' subscript, and those corresponding to negative D-axis voltage are shown in dashed lines, with a '-' subscript. The negative D-axis voltage case results in leading power factor at the stator terminals and the positive D-axis voltage case results in lagging power factor at the stator terminals. Since the overall magnetic circuit is inductive, the negative D-axis case leads to additional compensating inductive vars to be supplied from the rotor MSC. This is avoided through the choice of positive D-axis injection. In this arrangement, the SGSC and MSC together supply reactive power needs of the magnetic circuit of the DFIG.

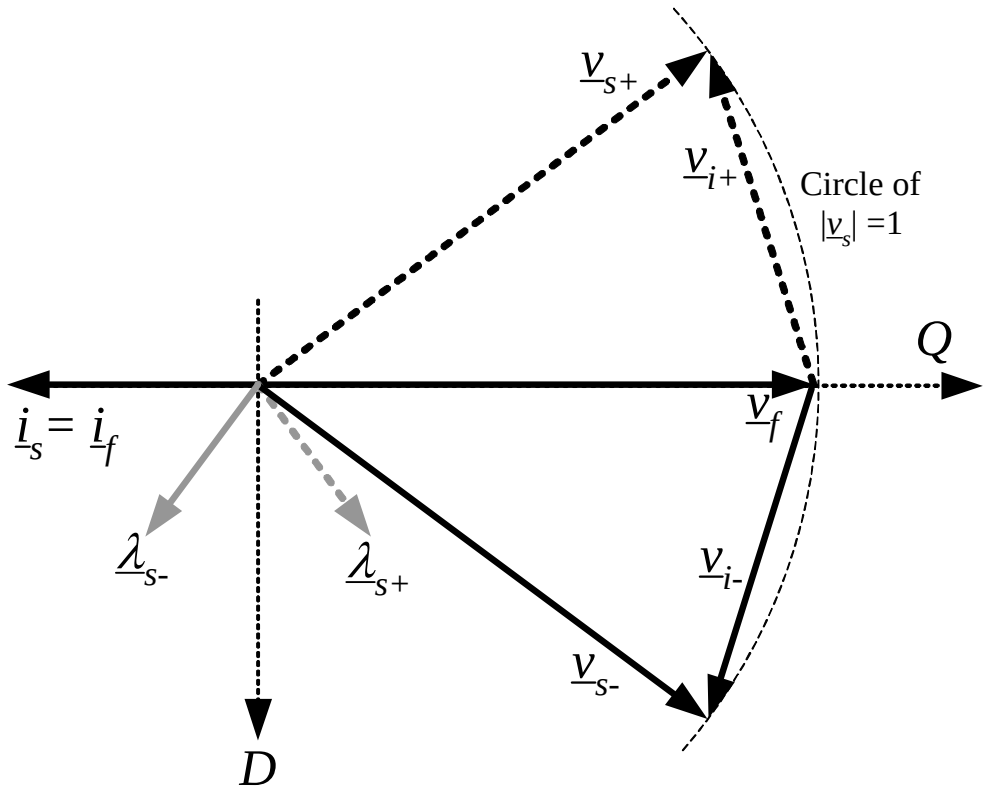

Fig. 9. Space vector diagram of stator, SGSC and farm collector voltages, and stator/farm collector current with d-axis SGSC & stator voltage to increase stator voltage magnitude

Under this operating condition, the reactive power flow into the stator terminals can be expressed in terms of the reactive power required to magnetize the machine, less that which is supplied by the MSC, as in

$$Q_s = \frac{3|\underline{\lambda}_s|\omega_e}{2}\left(\frac{|\underline{\lambda}_s|}{L_s}+i_{QR}^{\lambda s}\right). \tag{18}$$

The reactive power into the SGSC is

$$Q_i = \frac{3|\underline{i}_s|v_{Di}^{is}}{2}. \tag{19}$$

The net reactive power into the farm collector node is equal to the sum of the stator and SGSC reactive powers, as in

$$Q_f = Q_s + Q_i. \tag{20}$$

It may be seen that the reactive power injection may be regulated either by appropriately controlling $i_{QR}^{\lambda s}$ without disturbing primary torque production variables, namely $i_{DR}^{\lambda s}$ or $|\lambda_s|$.

During supersynchronous operation, the SGSC is used to transfer power from the dc link to the ac grid, and maintain regulation of the dc link voltage. By neglecting converter losses, the dc link voltage state equation (4) can be approximated as follows:

$$\frac{dv_{dc}}{dt} = \frac{-3}{2C_{dc}v_{dc}}\Big[|\underline{i}_s|\left(v_{Qf}^{is}+\omega_e|\underline{\lambda}_s|\sin(\theta_{\lambda s}-\theta_{is})\right) - \omega_{sl}|\underline{\lambda}_s|\, i_{DR}^{\lambda s} - |\underline{i}_R|^2 R_R\Big]. \tag{21}$$

It may be seen that the dc link voltage can be regulated by changing the angle of the stator flux with respect to the stator current via the term $\sin(\theta_{\lambda s}-\theta_{is})$ term without affecting the torque producing variables.

*ii) Subsynchronous operation*

A phasor diagram of the stator current and various voltages under UUPF operation at subsynchronous speeds is shown in Fig. 10. It may be observed that $v_{Qs}>v_f$, and $v_{Qi}>0$. Operating the machine under UUPF approach would lead to increased stator voltage beyond the nominal collection voltage, (and hence increased stator flux) and would potentially lead to magnetic saturation if the stator is not suitably designed.

The minimum stator flux in the machine is approximately

$$|\underline{\lambda}_s| \ge \lambda_{Ds} = \frac{v_{Qc}}{\omega_e(1-s)}. \tag{22}$$

In the more general non-UUPF operation, equation (22) provides a lower bound on the stator flux magnitude. It may be observed that the minimum stator flux increases above the nominal value significantly for subsynchronous speeds (s>0). Operation of the DFIG with stator flux greater than 1 per unit is undesirable as this leads to magnetic saturation, or would otherwise necessitate significant over-sizing of the machine. Therefore, operation of the SGSC under subsynchronous mode of operation is undesirable.

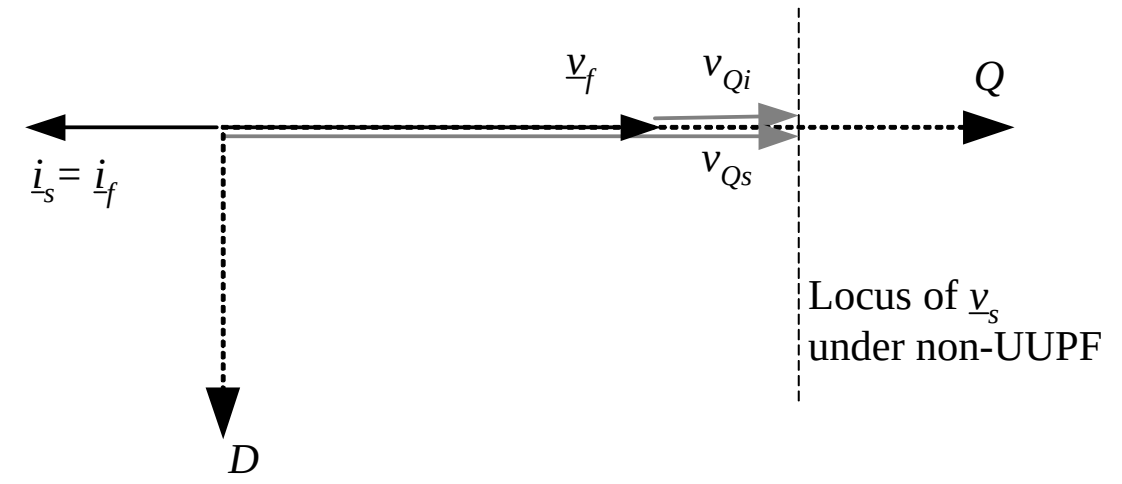

Fig. 10. Phasor diagrams of selected voltages and currents for unified DFIG architecture with only SGSC at subsynchronous speeds

However, since the rotor circuit receives power from the ac terminals of the MSC under subsynchronous operation, the PGSR readily takes on the role of maintaining dc link power balance. Thus, unified DFIG architecture presented in Fig. 2 maintains functionality over the complete range of operation. As can be seen from Fig. 7, the maximum power throughput of the PGSR is small, less than 0.1 per unit. Furthermore, since the PGSR is also a passive network realized using diodes, the additional converter rating introduced due its addition may be incremental.

## IV. Wind Turbine Control

The wind turbine control structure is hierarchical – global control objectives are achieved through manipulation of inputs to inner control loops. The dynamic and steady state operating characteristics are used to guide the division of the control responsibilities. Implementation details for each controller are explained in subsequent subsections.

### A. Global control structure

An illustration of the global control structure is presented in Fig. 11. Control objectives for the unified DFIG wind turbine architecture include regulation of the rotor speed, electrical power production, reactive power at the farm collector node, stator flux magnitude and the dc link voltage. Inner control loops on the MSC and SGSC regulate the rotor current and stator flux respectively. MSC current commands are generated from field oriented torque control and farm collector reactive power control loops. Commands for the SGSC stator flux regulator are generated from outer loop regulators for the dc link voltage and the stator flux magnitude. These two outer loops feeding the flux command are designed to accommodate the transition be sub- and super-synchronous operation and the handoff of power processing responsibilities between the SGSC and PGSR. The turbine rotor speed is regulated via blade pitch actuators to limit to reduce the coefficient of performance and throttle mechanical torque production. In addition to other feedback signals, measurement of the farm collector voltage enables dynamic response and ride through of voltage sags.

The regulation of various controlled quantities is implemented in coordinate references that are natural to realize the control in a natural manner. For instance, the rotor D-axis and Q-axis current commands are defined with the location of the stator flux $\underline{\lambda}_s$ as the reference direction for the Q-axis. This facilitates appropriate decoupling of controlled variables and control inputs.

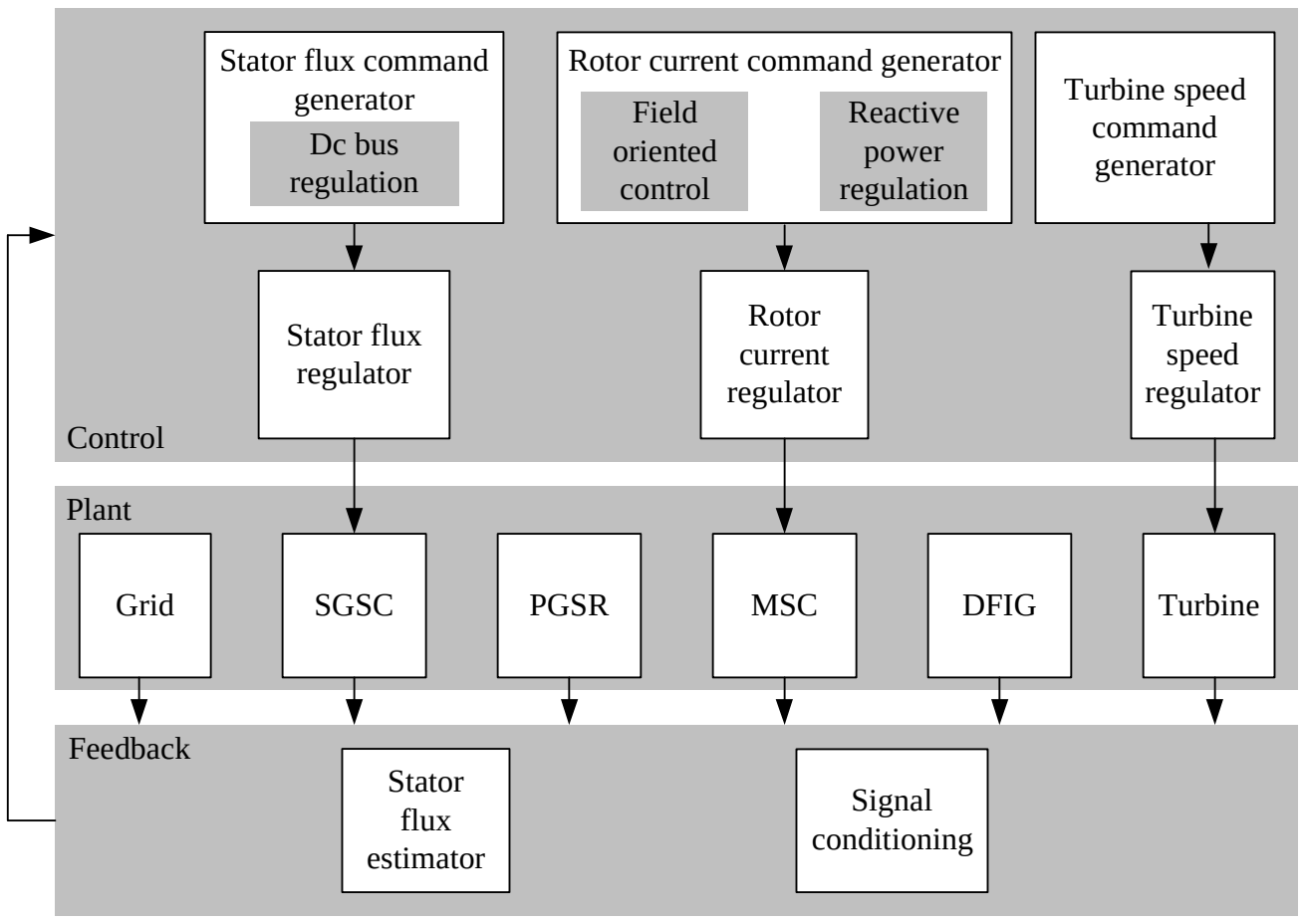

Fig. 11. Global control structure for unified DFIG wind turbine architecture

### B. Turbine speed command generator and regulator

The turbine rotor speed is regulated by throttling the mechanical torque production through pitching of the turbine blades. The commanded rotor speed is set to the maximum, typically 1.2 per unit. A saturation block prevents limits the rotor speed command to a safe level. The rotor speed error drives a proportional-integral regulator which generates the blade pitch angle command. Feedforward of voltage sag information may further be used to improve the blade pitch response for sags of long duration.

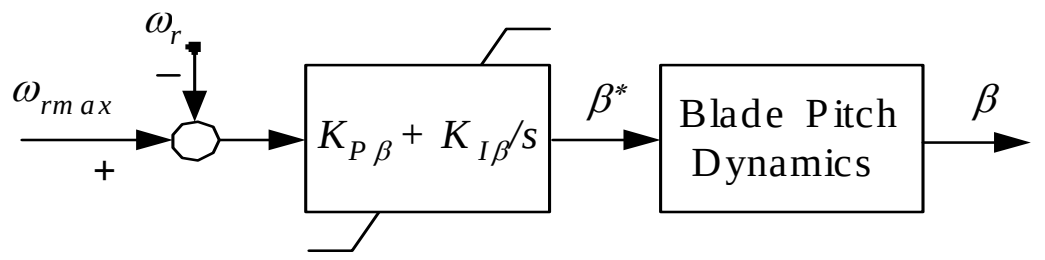

Fig. 12. Block diagram of turbine speed command generator and speed regulator

### D. Stator flux command generator

#### i) Stator flux magnitude

The overall stator flux command generator block diagram is presented in Fig. 13. The nominal stator flux magnitude command is scaled to the magnitude of the farm voltage, nominally equal to 1 per unit, specified as

$$\left|\underline{\lambda}_s\right|^* = \frac{\left|\underline{v}_f - \underline{i}_s \hat{R}_s\right|}{\omega_e} . \tag{23}$$

This allows the stator flux controller to respond quickly to abrupt changes in the farm collector voltage which accompany grid voltage sags. Stator flux magnitude scheduling may further be used during super-synchronous operation to improve system efficiency or achieve other operating benefits.

#### ii) Stator flux angle

As was described using (21), the angle between the stator current and the stator flux provides a natural control variable for regulating the DC link voltage. Feedback linearization technique is used to transforms the nonlinear relationship between the controllable input, which is the component of the stator flux orthogonal to the stator current $\lambda_{Ds}^{is}$, and the output to the system, $v_{dc}$, into a one in which the dynamic relationship between the error, $v_{dc} - v_{dc}^*$, and new input, $\xi$, is linear, exhibits a first order response. Application of the process of the feedback linearization process yields the following relationship for the commanded D-axis stator flux:

$$\underline{\lambda}_{Ds}^{is\,*} = \frac{1}{|\underline{i}_s|\omega_e}\left[\begin{array}{c}\dfrac{2\hat{C}_{dc} v_{dc} K_{Pvdc}(v_{dc}^{*} - v_{dc})}{3}\cdots \\ + v_{Qf}^{is}|\underline{i}_s| - \omega_e|\underline{\lambda}_s| i_{DR}^{\lambda s} + |\underline{i}_R|^2 \hat{R}_R\end{array}\right]. \qquad (24)$$

The proportional gain, $K_{Pvdc}$, has units of rad/s and is chosen to be 20 rad/s.

*D. Rotor current command generator*

*i) Field Oriented Control of Torque*

The electromagnetic torque is controllable via the component of the rotor current orthogonal to the stator flux. The reference for the electromagnetic torque command is produced from a lookup table using the rotor speed as the input as is common for traditional DFIG wind turbine architectures [13]. The torque-speed command profile follows the optimum tip-speed-ratio at low and medium rotor speeds. At high rotor speeds, typically around 1.2 per unit, the torque increases linearly, and limited to rated torque. The torque command is divided by the nominal stator flux and number of pole pair to produce the orthogonal rotor current command. In the event of a grid voltage sag, the rotor current command remains roughly constant and the resulting electromagnetic torque and power production scales naturally with the stator flux magnitude.

*ii) Reactive power control*

A block diagram of the overall stator flux command generator that includes the reactive power controller is shown in Fig. 14. The reactive power at the farm collector node is controlled the component of the rotor current that is collinear with the stator flux, $i_{QR}^{\lambda s}$. From equations (19) and (20), an algebraic relationship is developed for the $i_{QR}^{\lambda s*}$ from measured quantities as

$$i_{QR}^{\lambda s\,*} = \frac{-2}{3|\lambda_s|\omega_e}(Q_f^{*} + Q_i) + \frac{|\lambda_s|}{\hat{L}_s}. \qquad (25)$$

The reactive power command ($Q_f^*$) is nominally chosen to be zero in the paper, although other values may be realized to satisfy additional operating requirements.

*E. Rotor current regulator*

Control of the rotor current is achieved via a high bandwidth synchronous frame proportional loop which generates modulation functions for the MSC, illustrated in Fig. 15. The difference between the commanded and measured rotor current is multiplied by proportional gain.

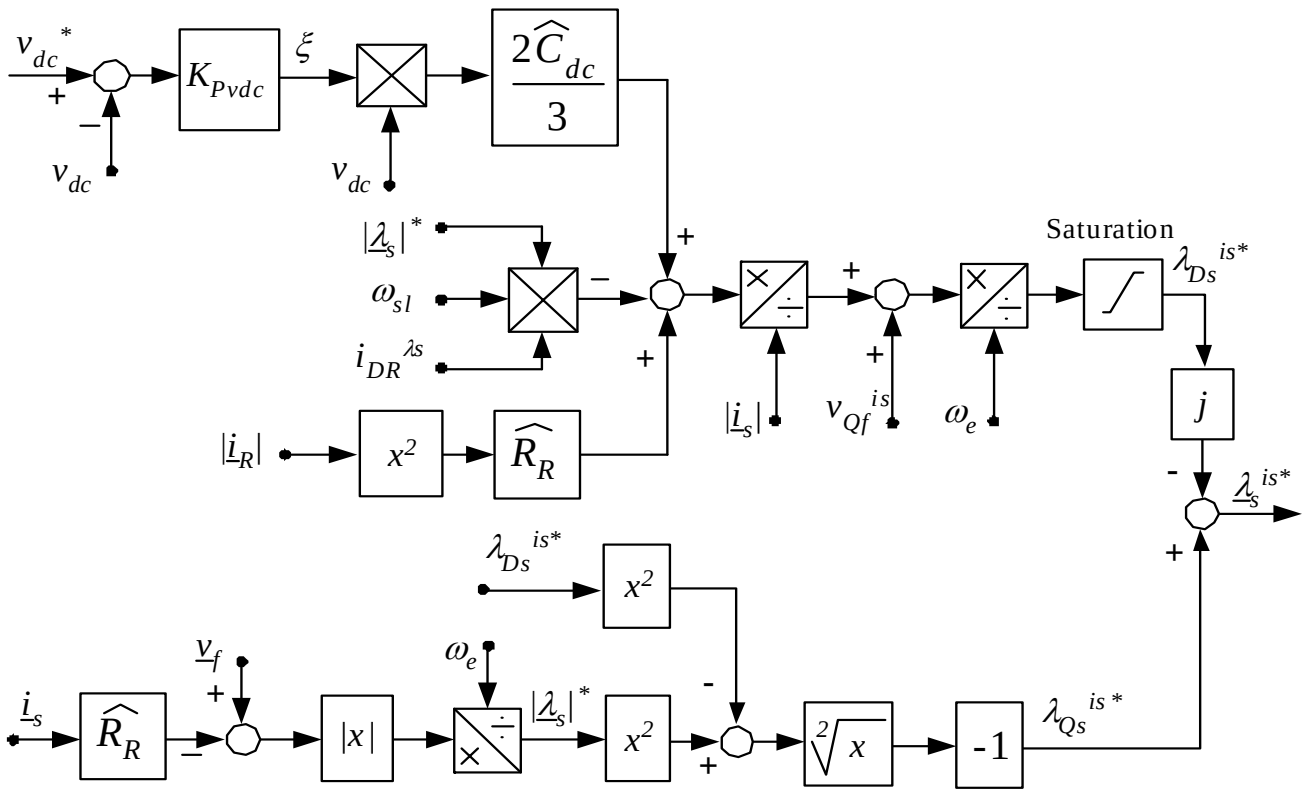

Fig. 13. Block diagram of stator flux command generator

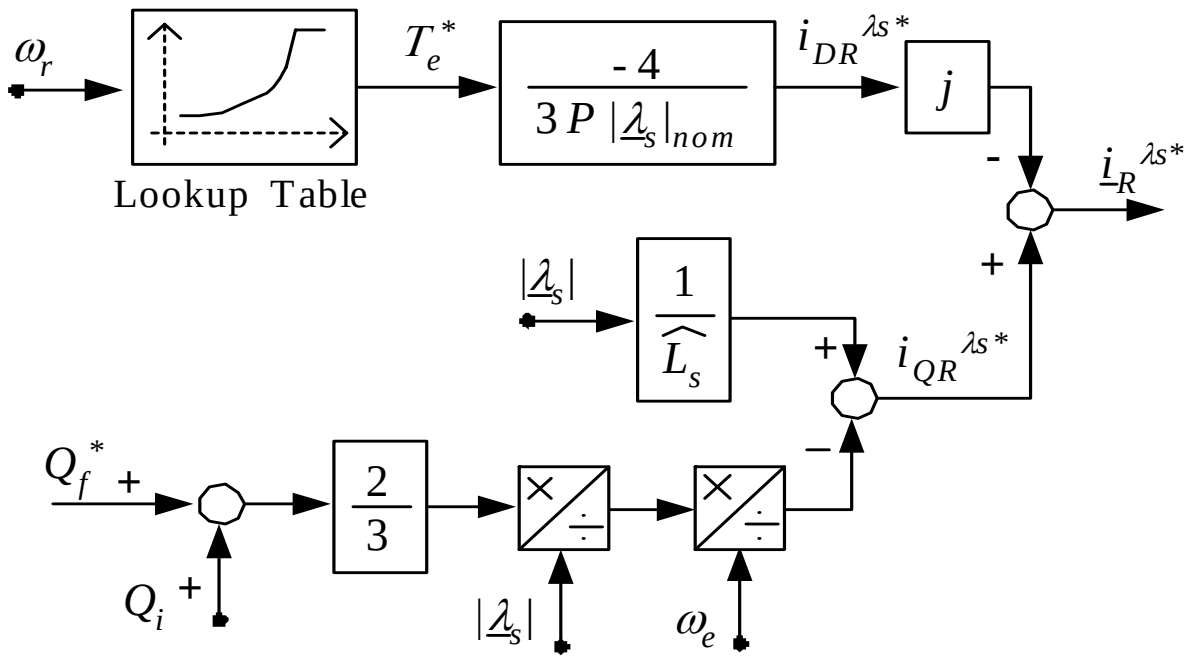

Fig. 14. Block diagram of rotor current command generator

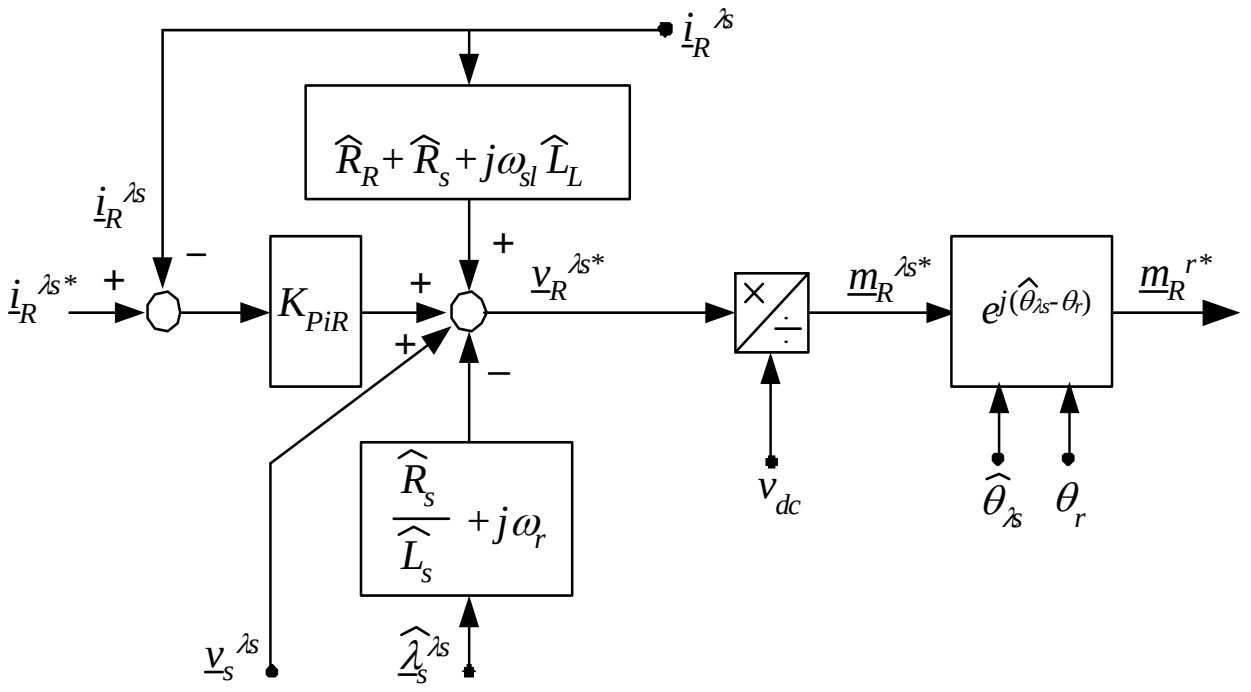

Fig. 15. Block diagram of the rotor current regulator

Estimates of machine parameters, measurements of rotor current, stator and dc link voltage, rotor and slip speeds, and an estimate of the stator flux are further used to improve steady state and dynamic tracking accuracy and disturbance rejection [14]. The proportional gain, $K_{PiR}$, has units of $\Omega$, and is chosen to be 3 per unit to yield a bandwidth of 900 Hz.

The regulator produces the complex modulation function space vector for the MSC in the synchronous frame. Individual scalar IGBT modulation functions are determined by abc decomposition of the complex modulation space vector in the reference frame aligned with the rotor, using appropriate pulse width modulation algorithms.

*F. Stator flux regulator*

The stator flux regulator illustrated in Fig. 16 is the key feature of the proposed architecture that transforms the undesirable oscillatory stator flux response during faults into a well behaved and bounded exponential dynamic response.

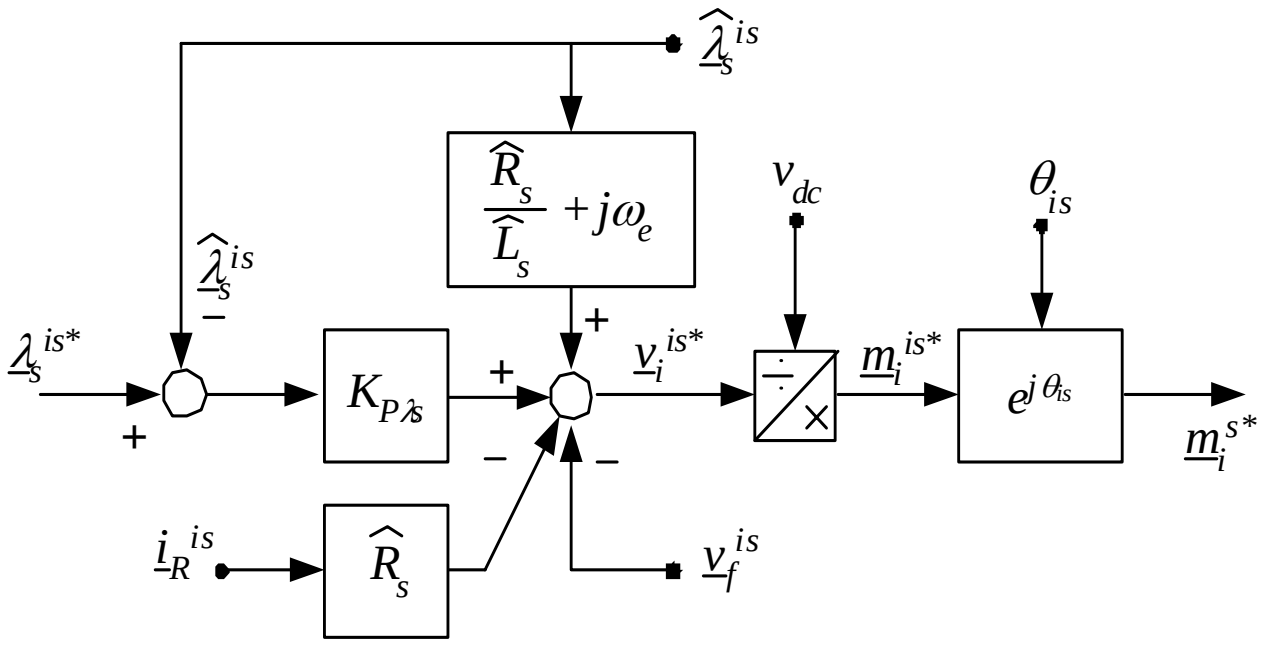

Fig. 16. Block diagram of stator flux regulator

Similar to the rotor current regulator, control of the stator flux is achieved via a synchronous frame proportional loop which generates modulation functions for the SGSC. The difference between the commanded and estimated stator flux is multiplied by proportional gain. Estimates of machine parameters, measurements of rotor current, farm collector node and dc link voltage, excitation frequency, and an estimate of stator flux are further used to improve steady state and dynamic tracking accuracy and disturbance rejection. The proportional gain, $K_{p\lambda s}$, has units of rad/s, and is chosen to be 94 rad/s. The regulator output is the complex modulation function space vector for the SGSC in the synchronous frame. Individual scalar IGBT modulation functions are determined by abc decomposition of the stationary frame complex modulation space vector using appropriate pulse width modulation algorithms.

### G. Stator flux estimator

Besides various feedback signals of currents and voltages as indicated in the block diagrams, it is necessary to obtain a reliable measurement of the stator flux for field oriented control as well as overall system operation. Various stator flux estimation techniques may be used for this purpose. Since the DFIG is always exited at the synchronous frequency $\omega_e$, stator flux estimation can be made using a simple approach using the integral form of (1) in the stationary reference frame, (designated by a double underscore under the complex vectors) [15] with a high degree of accuracy.

$$\underline{\underline{\lambda}}_s = \int \left( \underline{\underline{v}}_s + \hat{R}_s \underline{\underline{i}}_s \right) dt \qquad (26)$$

Various stationary reference frame signals including the currents, voltages and flux estimates are transformed into the reference coordinates for implementing the feedback control using appropriate transformations [10].

## V. Dynamic Response

A detailed computer simulation model of an example system was developed to study the operation of the system. The model included all the machine and controller details presented in the previous sections. A list of numerical values of machine parameters used in the simulation is presented in the Appendix. Selected waveforms from the simulations of the system are presented in this section. The response of the system to changes in wind speed, showing transition from supersynchronous to subsynchronous operation and vice-versa is presented in Fig. 17. As may be observed, the controller successfully manages the transition between sub and super-synchronous operation in a smooth manner, particularly handing over the dc link power balance between SGSC and PGSR. The response of the unified DFIG wind turbine architecture to voltage sag is presented in Fig. 18.

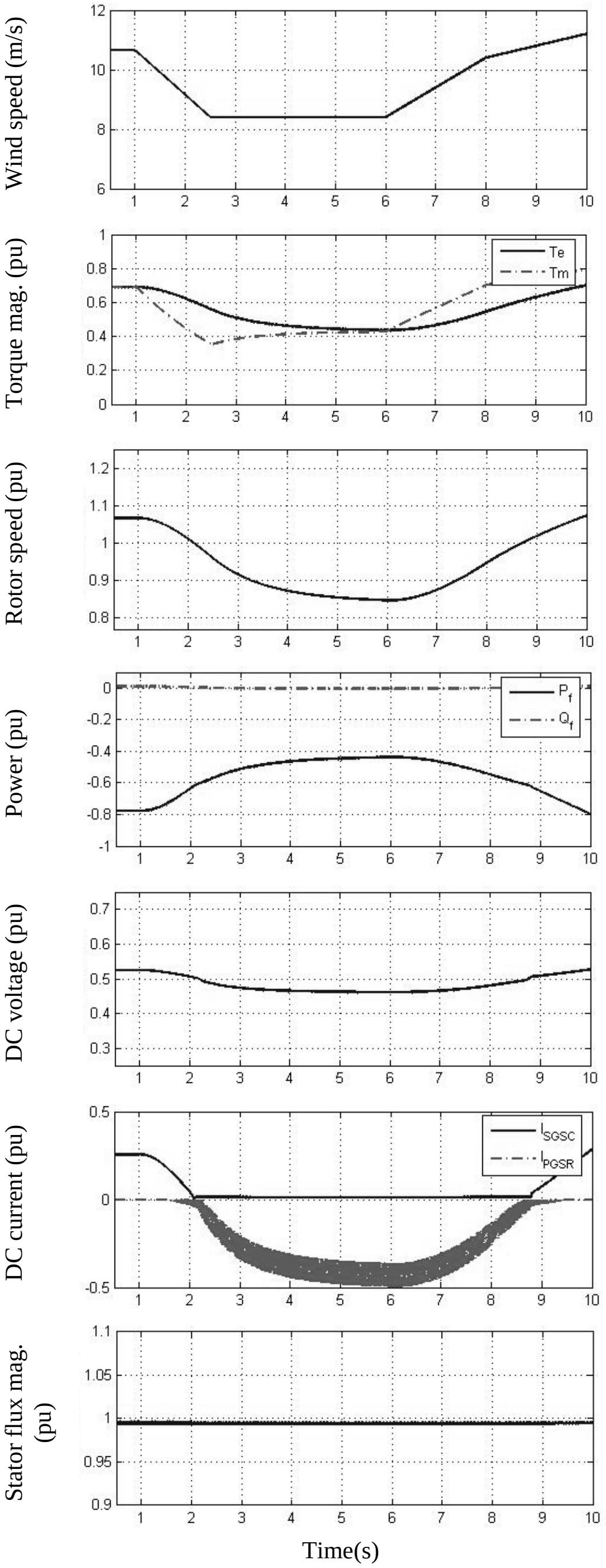

Fig. 17. Simulated response of a 2 MW unified architecture DFIG wind turbine to ramp changes in wind speed.

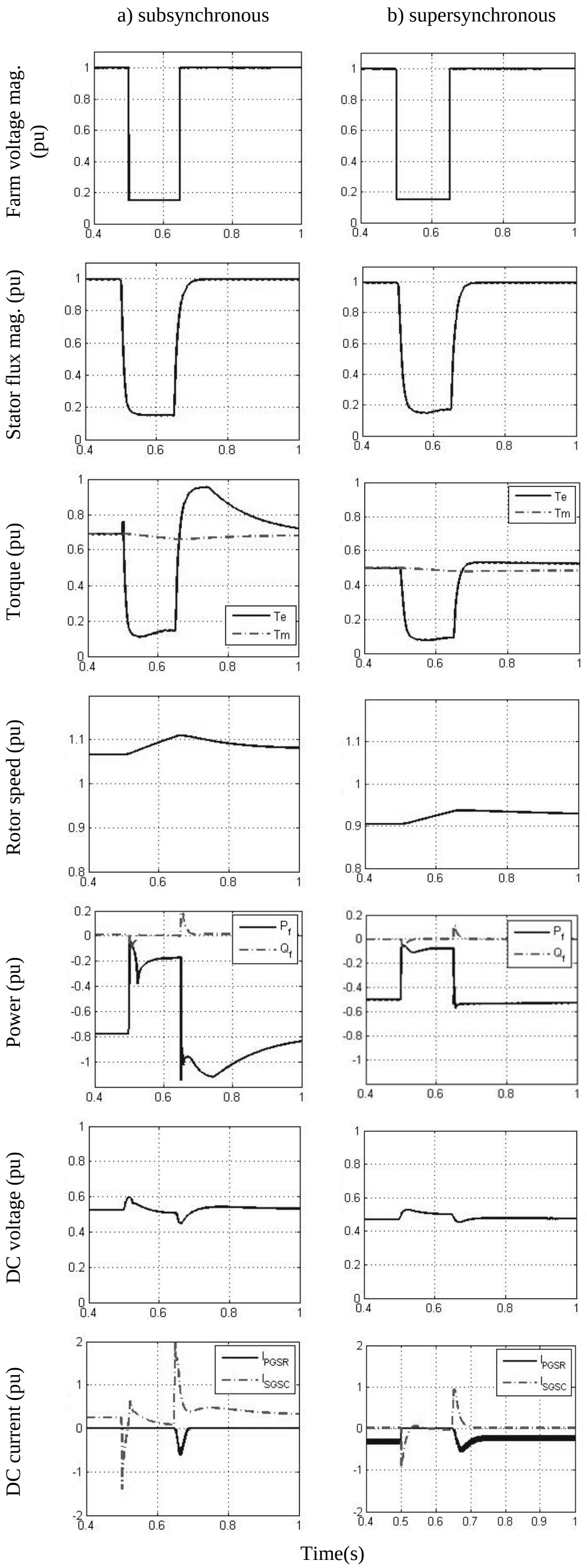

Fig. 18. Simulated response of a 2 MW unified architecture DFIG wind turbine to farm voltage sag occurring at t =0.5 sec, during: a)sub-synchronous; and b)super-synchronous operation.

A voltage sag that drops the terminal voltage at the wind farm collection node to 0.15 p.u. and lasts for 150 ms (9 cycles) is applied at the wind farm collection terminal voltage $\underline{v}_f$. The traces on Fig. 18(a) and (b) represent the response during subsynchronous and supersynchronous operating speeds respectively. As may be observed from the figure, the stator flux features smooth and bounded transients during the sag event. All other key variables also exhibit a well behaved response within the design boundaries of the system.

## VI. Conclusions

Provision of adequate ride-through capability under utility fault conditions has been known to be a challenging problem for wind generation systems that use doubly fed induction machines with dc link power converters in the rotor circuit. This paper has briefly reviewed existing techniques for meeting the challenge, and introduced an improved unified architecture that is inspired by power quality improvement devices such as the dynamic voltage restorer.

The unified architecture maintains an active machine side power converter to recover the rotor slip power into a dc link as in the traditional DFIG system. However, a series-connected grid side converter is used to inject the dc link power into the grid. This enables machine operation with tight regulation of stator flux during normal conditions and during voltage sags. Although this approach is capable of controlled power flow over a typical operating range above and below synchronous speed, it is shown that machine utilization suffers at subsynchronous speeds. Therefore, a parallel connected passive rectifier rated at a small fraction of the total power throughput is used to restore maximal utilization of overall system. A detailed dynamic electrical model for the system along with solutions for steady state operation is presented in the paper to elucidate these aspects.

A scheme for control of the DFIG system that incorporates field oriented control to regulate torque, stator flux and dc link voltage is presented. The proposed controller features unity power factor at the injection terminals, although reactive power injection can be introduced through appropriate command inputs. Design details of various feedback regulators are presented in the paper. A computer simulation of the entire system is used to verify the performance of the system.

Waveforms of selected quantities from the computer simulation that demonstrate dynamic performance during transitions between subsynchronous and supersynchronous speeds in response to changes in wind speed are presented. Successful ride-through performance of the system during a deep voltage sag event is also demonstrated using computer simulations. The computer simulation results indicate excellent ride-through performance with smooth and well bounded transient in comparison to existing solutions. Moreover, the series connected grid side converter is also expected to enable operation during single phase faults with relative ease even when compared to double conversion systems, as has been demonstrated in dynamic voltage restorer systems that feature a similar architecture.

The performance capabilities of the proposed unified architecture may be extended further by realizing the parallel grid side rectifier using an active rectifier instead of the passive approach presented in this paper. Detailed design trade-offs that determine the structure and actual ratings of the power converters and detailed dynamic analysis of the system providing system stability margins, etc, are the subject of continuing investigations, and will be presented in future publications.

## Acknowledgments

The authors would like to thank the sponsors of the Wisconsin Electric Machines and Power Electronics Consortium (www.wempec.org) at the University of Wisconsin – Madison and the Link Foundation Energy Fellowship Program (www.linkfoundation.org) for their support. This work made use of ERC Shared Facilities supported by the National Science Foundation under Award Number EEC-9731677.

## Appendix

### List of Symbols and Parameters

| Symbol | Parameter | Value |
|---|---|---|
| $j$ | Imaginary operator | $\sqrt{-1}$ |
| $\cdot$ | Space vector dot product | |
| $\times$ | Space vector cross product | |
| $\hat{}$ | Estimated quantity | |
| $\underline{\ }$ | Complex quantity | |
| $\omega_e$ | Synchronous speed | 2π 60 |
| $\omega_r$ | Electrical rotor speed | 2π 40 to 2π 80 |
| $\omega_{sl} = \omega_e - \omega_r$ | Electrical rotor slip speed | +/- 2π 20 |
| $s = \omega_{sl} / \omega_e$ | Slip | +/- 0.333 |
| $L_m$ | Magnetizing inductance (stator referred) | 19.5 mH |
| $L_{ls}$, $L_{lr}$ | T model stator and rotor leakage inductance (stator referred) | 0.86, 0.55 mH |
| $R_s$, $R_r$ | Stator and rotor resistance (stator referred) | 6.2, 6.2 mΩ |
| $C_{dc}$ | DC link capacitance | 90 mF |
| $J_m$ | Total rotational inertia (referred to high speed side of gearbox) | 105 Nm/s$^2$ |
| $P$ | Number of DFIG pole pairs | 4 |
| $u_w$ | Wind speed | 12 m/s rated |
| $\psi$ | Tip speed ratio | 6.9 |
| $\beta$ | Blade pitch angle | 0 to 30 deg |
| $C_P$ | Coefficient of performance | 0.47 max |
| $R_b$ | Rotor blade diameter | 37.5 m |
| $L_s = L_m + L_{ls}$ | Stator inductance (stator referred) | 20.4 mH |
| $L_r = L_m + L_{lr}$ | Rotor inductance (stator referred) | 20.0 mH |
| $N_{rs}$ | Rotor to stator turns ratio | 0.733 |
| $\gamma = L_m/L_s$ | Ratio of magnetizing to stator inductance | 0.958 |
| $L_L=L_{lr}/\gamma^2+L_{ls}/\gamma$ | Γ model leakage inductance | 1.5 mH |
| $R_R=R_r/\gamma^2$ | Γ model rotor resistance | 6.7 mΩ |
| $\underline{v}_f$ | Wind farm voltage space vector | 2 kV pk |
| $\underline{v}_s$ | Stator voltage space vector | 2 kV pk |
| $\underline{v}_r$ | Rotor voltage space vector | 730 V pk |
| $v_{dc}$ | DC link voltage | 1000 V nom |
| $\underline{i}_s$ | Stator current space vector | 673 A pk |
| $\underline{i}_r$ | Rotor current space vector | |
| $\underline{i}_{re}$ | PGSR current space vector | |
| $\underline{m}_i$ | SGSC modulation space vector | |
| $\underline{m}_R$ | Γ model MSC modulation space vector | |
| $\underline{m}_{re}$ | PGSR modulation space vector | |
| $\underline{v}_R=\underline{v}_r/\gamma$ | Γ model rotor voltage space vector | |
| $\underline{i}_R= \gamma\,\underline{i}_r$ | Γ model rotor current space vector | |
| $\underline{\lambda}_s = L_s\left(\underline{i}_s + \underline{i}_R\right)$ | Stator flux space vector | |
| $\underline{\lambda}_R = \underline{\lambda}_s + L_L\,\underline{i}_R$ | Γ model rotor flux space vector | |
| $T_e$ | Electromagnetic torque | 10 kNm pk |
| $T_m$ | Mechanical torque referred to rotor speed | 10 kNm pk |
| $P_m$ | Mechanical Power | 2.1 MW pk |